\documentstyle[epsfig,amssymb,aps,pra]{revtex}

\def\ket#1{|#1\rangle}
\def\bra#1{\langle#1|}
\def\braket#1#2{\langle#1|#2\rangle}
\def\sandwich#1#2#3{\langle#1|#2|#3\rangle}
\def\be{\begin{equation}}
\def\ee{\end{equation}}

\begin{document}

\draft
\narrowtext
%\widetext

\wideabs{
\title{Enhanced Quantum Estimation via Purification}

\author{Holger Mack, Dietmar G. Fischer, and Matthias Freyberger}

\address{Abteilung f\"ur Quantenphysik, Universit\"at Ulm, D-89069 Ulm, Germany} %
\maketitle

%--------------------------------------------------------------------------------
\begin{abstract}
We analyze the estimation of a finite ensemble of quantum bits which have been sent through a depolarizing channel. Instead of using the depolarized qubits directly, we first apply a purification step and show that this improves the fidelity of subsequent quantum estimation. Even though we lose some qubits of our finite ensemble the information is concentrated in the remaining purified ones.
\end{abstract}


}

\narrowtext

%--------------------------------------------------------------------------------
\section{Introduction}
\label{introduction}

An essential step of information processing, either classical or quantum, is the read-out of data. Reading a classical bit does not pose fundamental problems. A single measurement suffices to reveal the content $0$ or $1$ of a classical memory cell. However, already in the classical case this information will be lost eventually due to dissipative processes. In fact each classical dynamic computer memory requires permanent cycles of reading and refreshing in order to prevent this dissipation of data.

This situation changes considerably in the quantum domain. As soon as a classical apparatus registers a signature of a quantum system its original quantum state changes. This measurement process is effectively formulated by the von-Neumann rules \cite{vonNeumann}. In the following we will accept this description of a quantum measurement rather than questioning its philosophical implications. Neither will we try to derive the von-Neumann process using quantum dynamics. For the present article we take this specific quantum feature of the measurement process for granted and consider its impact on quantum information processing.

As a consequence of the von-Neumann description it is impossible to unseal the complete content of a single quantum bit in form of the corresponding quantum state. This has lead to investigations whether the state of a quantum bit can be at least optimally estimated \cite{HolevoHelstrom} from a finite number of identical copies. The intriguing result of this research has been that optimal estimations cannot be realized by single measurements on each qubit separately. Rather one has to see the finite ensemble as one composite system which allows us to perform a generalized measurement \cite{pureEst,mixedEst}.

However, the realization of such measurements is a difficult task. First, we might get the identical copies of our qubit sequentially in certain situations. Just imagine the situation of ``quantum debugging'', where a quantum algorithm is run several times on a quantum computer. In this case, the sequential estimation of a specific qubit could be used as a test for the perfect performance of the algorithm. Second, the physical implementation of the generalized measurement remains an open problem so far.

Hence one has to ask how close single measurements can come the optimal ones \cite{Gill,Fischer}, in particular, if we combine them with a strategy. That is, if we learn from each measurement step and adapt the next one appropriately. Such schemes have been discussed for pure qubits in Ref. \cite{Fischer}. But pure qubits are rather ideal. As soon as they are connected to an environment they will start to decohere and gradually lose their original information content. Hence the original pure state will turn into a mixed quantum state depending on the decoherence model \cite{Decoherence}. 

Is it possible to refresh such decohered qubits as in the classical case? One answer to this is quantum error correction \cite{ErrorCorrection}. But it is also possible to purify a finite ensemble of qubits if we sacrifice some of them. This has been shown in Ref. \cite{Cirac}: The remaining qubits after the purification step have indeed a higher single-qubit fidelity than the unpurified ones.

In the present paper we shall investigate if such a purification step also allows us to retrieve more information about the originally pure state of the qubits. Does it make sense to spend some of the decohered qubits for the purification step and to estimate the underlying pure state now from a smaller ensemble using single adaptive measurements? What will be the influence of the entanglement between the purified qubits which is an unavoidable consequence of the purification protocol presented in Ref. \cite{Cirac}. These questions lie at the center of the present paper.

In Section \ref{purification} we recall the basic elements of single qubit purification \cite{Cirac} needed for our purpose. We then combine this with an adaptive estimation strategy in Section \ref{estimation}. Our results are discussed in Section \ref{results} and we conclude with Section \ref{conclusions}.

%--------------------------------------------------------------------------------
\section{Purification of qubits}
\label{purification}

In this section we briefly review the single qubit purification proposed in Ref.\ \cite{Cirac} and introduce our notation. We assume that Alice has a source of single qubits in the pure state
\be
  \ket1_{\vec n}\equiv\cos\frac\Theta2\ket1+\sin\frac\Theta2e^{i\Phi}\ket0
\ee
defined by the basis states $\{\ket0,\ket1\}$ and the point $\vec n=\left(\sin\Theta\cos\Phi,\sin\Theta\sin\Phi,\cos\Theta\right)$ on the Bloch sphere. These qubits are sent to Bob via a depolarizing channel and hence Bob receives qubits described by a mixed state
\be 
  \rho=c_1\ket1_{\vec n}\bra1+c_0\ket0_{\vec n}\bra0,
  \label{origrho}
\ee
in which $\ket1_{\vec n}$ occurs with probability \cite{Footnote1} $c_1\geq 1/2$ and the corresponding orthogonal state
\be 
  \ket0_{\vec n}\equiv\cos\frac\Theta2\ket0-\sin\frac\Theta2e^{-i\Phi}\ket1 
\ee
with probability $c_0$, so that $c_0+c_1=1$.

If Alice sends an even number \cite{Footnote2} of $N$ qubits to Bob, the corresponding product state reads
\be 
  \rho^{\otimes N}\equiv\rho\otimes\rho\otimes\dots\otimes\rho. 
  \label{productrho}
\ee
Note that the qubits are assumed to be distinguishable.

The optimal purification protocol proposed in Ref.\ \cite{Cirac} shows how one can distill $M\leq N$ purer qubits from $\rho^{\otimes N}$ by a measurement performed on the $N$ qubits. This measurement projects the state $\rho^{\otimes N}$ on the output state
\be
  \rho_{out}=\rho_M\otimes\left(\ket{\Psi^-}\bra{\Psi^-}\right)^{\otimes(N-M)/2}
  \label{rhoout}
\ee
with probability
\begin{eqnarray}
  p_M&=&\left[{N\choose\frac{(N-M)}2}-{N\choose\frac{(N-M)}2-1}\right] \nonumber\\
  &&\times\left[c_0c_1\right]^{(N-M)/2}\frac{c_1^{M+1}-c_0^{M+1}}{c_1-c_0}
  \label{probm}
\end{eqnarray}
for values $M=0,2,\dots,N$. Hence one obtains $M$ qubits in the entangled state
\begin{eqnarray}
  \rho_M&=&\frac{c_1-c_0}{c_1^{M+1}-c_0^{M+1}}(M+1) \nonumber\\
  &&\times\int\frac{d\Omega}{4\pi}n(\theta)^{M} \left(\ket{\Psi(\theta,\phi)}\bra{\Psi(\theta,\phi)}\right)^{\otimes M} 
  \label{purirho}
\end{eqnarray}
with
\be 
  n(\theta)=c_1\cos^2(\theta/2)+c_0\sin^2(\theta/2) 
\ee
and the normalized state
\be 
  \ket{\Psi(\theta,\phi)}=\sqrt{c_1}\frac{\cos(\theta/2)}{\sqrt{n(\theta)}}\ket1_{\vec n}+\sqrt{c_0}\frac{\sin(\theta/2)}{\sqrt{n(\theta)}}e^{i\phi}\ket0_{\vec n}. 
\ee
The remaining $N-M$ qubits are maximally entangled in pairs of Bell states $\ket{\Psi^-}=\left(\ket1\ket0-\ket0\ket1\right)/\sqrt2$ and therefore carry no information about the original qubit.

The $M$ qubits in state $\rho_M$ are indeed purer than the original qubits described by $\rho$, Eq.\ (\ref{origrho}). That is, the reduced density operator ${\rm Tr}_{M-1}(\rho_M)$ obtained by tracing over $M-1$ qubits lies closer to the pure state $\ket1_{\vec n}$ than $\rho$, which means that the single qubit fidelity
\begin{eqnarray}
  f_M&=&\ \langle_{\!\!\!\!\!\!\!\vec n\,\,\,\,}1|{\rm Tr}_{M-1}(\rho_M)|1\rangle_{\vec n}\nonumber\\
  &=&\frac1M\left[\frac{(M+1)c_1^{M+1}}{c_1^{M+1}-c_0^{M+1}}-\frac{c_1}{c_1-c_0}\right] 
  \label{purifid}
\end{eqnarray}
for $M>0$ fulfills $f_M\geq c_1$. Note that $c_1=\ \langle_{\!\!\!\!\!\!\!\vec n\,\,\,\,}1|\rho|1\rangle_{\vec n}$ represents the fidelity of the original mixed state $\rho$, Eq.\ (\ref{origrho}). For the case $M=0$ the output state Eq.\ (\ref{rhoout}) consists only of $\ket{\Psi^-}$ states which carry no information about the original qubit, i.e., {$f_0=1/2$}. However, for the mean fidelity we still find
\be 
  \sum\limits_{M=0\atop\mbox{\tiny(even)}}^{N}p_Mf_M\geq c_1
\ee
which demonstrates purification.

We emphasize, however, that the single qubit fidelity $f_M$ does not yet quantify the effects of a specific measurement sequence performed on the purified qubits. In fact they are entangled and a measurement on one qubit will change the state of all others. This point of view will be important for the next section.

%--------------------------------------------------------------------------------
\section{Purification and estimation}
\label{estimation}

We will now look at this purification protocol from a different perspective. So far it was shown in Ref.\ \cite{Cirac} that a purified qubit lies closer to the original qubit prepared in the state $\ket1_{\vec n}$. However, the question arises whether the $M$ qubits in the state $\rho_M$ are also purer in the sense that they allow us to extract more information about the state $\ket1_{\vec n}$ than the $N$ unpurified qubits in state $\rho^{\otimes N}$.

The scenario we have in mind is the following. Alice prepares $N$ pure qubits and sends them to Bob via a depolarizing channel. Hence Bob obtains $N$ mixed qubits as described by Eq.\ (\ref{origrho}). He now has two possibilities to extract the information originally encoded by Alice. First, he can simply use the $N$ mixed qubits which are not correlated and estimate the pure qubit prepared by Alice. Second, he sacrifices some of the mixed qubits for a purification procedure as described in Section \ref{purification}. Hence after purification Bob obtains $M\leq N$ entangled qubits with probability $p_M$, Eq.\ (\ref{probm}). These qubits have an improved single-qubit fidelity $f_M$, Eq.\ (\ref{purifid}). 

Our question is whether the purified qubits will also allow us to improve the estimation of Alice's original qubits. It is important to note that our estimation only involves single qubit measurements. We will perform $M$ single measurements on the entangled state $\rho_M$, Eq.\ (\ref{purirho}). This means that the measurement of one qubit affects the state of the remaining qubits via the von-Neumann projection postulate. Despite this fact we shall show that it is possible to improve the estimation of Alice's qubit if we combine the purification protocol with an adaptive measuring scheme \cite{Fischer} for single qubits.

%--------------------------------------------------------------------------------
\subsection{Measurements on single qubits}
A single measurement will give us only little information about the corresponding qubit. But if we not only have one qubit we can proceed by measuring further qubits thus receiving more and more information about their quantum state. If we do this using an adaptive strategy for choosing the measurement directions we can give a good estimate of the underlying quantum state \cite{Fischer}.

This adaptive estimation strategy was originally formulated for uncorrelated qubits. However, the method does work for purified qubits as well. The only difference in this case is that the purification output or estimation input state is a non-separable $M$-qubit state. Therefore, the state of the qubits depends on the history of previous measurements on the system. One also has to take into account that the purification procedure is a probabilistic process. Hence purification results in several possible entangled states which must be weighted by the correct probabilities.

Let us start with the simpler case. Bob decides not to purify the qubits received but rather performs measurements on the finite ensemble described by $\rho^{\otimes N}$, Eq.\ (\ref{productrho}). Hence the qubits are uncorrelated and the $n$th polarization or spin measurement in direction $(\theta_n,\phi_n)$ is defined by the projector $\ket{\theta_n,\phi_n}\bra{\theta_n,\phi_n}$ with
\be
  \ket{\theta_n,\phi_n}\equiv\cos\frac{\theta_n}2\ket1+\sin\frac{\theta_n}2e^{i\phi_n}\ket0.
  \label{mdirection}
\ee
For each of the $N$ qubits he finds the result $1$ \cite{Footnote3} with probability
\be 
  p_1(\theta_n,\phi_n)=\sandwich{\theta_n,\phi_n}{\rho}{\theta_n,\phi_n}
\ee
and the result $0$ with probability
\be 
  p_0(\theta_n,\phi_n)=\sandwich{\overline{\theta_n,\phi_n}}{\rho}{\overline{\theta_n,\phi_n}}=1-p_1(\theta_n,\phi_n)
\ee
using the state
\be   \ket{\overline{\theta_n,\phi_n}}\equiv\cos\frac{\theta_n}2\ket0-\sin\frac{\theta_n}2e^{-i\phi_n}\ket1
\ee
orthogonal to $\ket{\theta_n,\phi_n}$, Eq.\ (\ref{mdirection}). After $n$ measurements the density operator of the remaining qubits reads $\rho^{\otimes(N-n)}$.

This simple situation changes considerably if Bob purifies the $N$ qubits before he starts measuring them. The purification procedure projects $\rho^{\otimes N}$ on the output state $\rho_{out}$, Eq.\ (\ref{rhoout}). With probability $p_M$, Eq.\ (\ref{probm}), he has lost $N-M$ qubits since they are now prepared in $\ket{\Psi^-}$ Bell states which carry no information about the original qubit $\ket1_{\vec n}$. This information is now concentrated in $M$ entangled qubits whose quantum state $\rho_M$, Eq. (\ref{purirho}), has a rather complicated structure. In particular, due to the entanglement a measurement performed on one qubit now influences the quantum state of the rest.

In the $n$th measurement Bob finds the result $1$ with probability
\be
  \tilde p_1(\theta_n,\phi_n)={\rm Tr}\Bigl[\ket{\theta_n,\phi_n}\bra{\theta_n,\phi_n}\tilde\rho_{M-n+1}\Bigr]
\ee
where the trace has to be taken over all $M-n+1$ qubits described by the conditioned density operator $\tilde\rho_{M-n+1}$. Conditioned here means that finding the result $1$ leads to the normalized density operator 
\be
  \tilde\rho_{M-n}=\frac{\sandwich{\theta_n,\phi_n}{\tilde\rho_{M-n+1}}{\theta_n,\phi_n}}{\tilde p_1(\theta_n,\phi_n)}
\ee
for the remaining $M-n$ qubits. Analogously, the result $0$ occurs with probability
\be
  \tilde p_0(\theta_n,\phi_n)=1-\tilde p_1(\theta_n,\phi_n)
\ee
and the corresponding reduced density operator reads
\be
  \tilde\rho_{M-n}=\frac{\sandwich{\overline{\theta_n,\phi_n}}{\tilde\rho_{M-n+1}}{\overline{\theta_n,\phi_n}}}{\tilde p_0(\theta_n,\phi_n)}.
\ee
Hence we clearly see that the quantum state $\tilde\rho_{M-n}$ after $n$ measurements strongly depends on the results ($1$ or $0$) obtained for the chosen measurement directions $(\theta_1,\phi_1)$, \dots, $(\theta_n,\phi_n)$. In the notation for the conditioned density operator $\tilde\rho_{M-n}$ we should therefore include these measurement directions and the corresponding results. But we leave this out for clarity of notation. However, both situations --- (i) the separable case for qubits not purified as well as (ii) the non-separable case for purified qubits --- can be combined with an adaptive measurement scheme.

%--------------------------------------------------------------------------------
\subsection{Adaptive measurements}
In the last paragraph we have described how the measurements work and how they change the quantum state of all $N$ qubits. However, we have not yet discussed how to choose the measurement direction $(\theta_n,\phi_n)$. In order to get an optimized qubit estimation we will apply an adaptive procedure \cite{Fischer}. That is, we will keep track of the information obtained from measurements $1$, $2$, \dots, $n-1$ in order to design the $n$th measurement. Various realizations of such adaptive measurements have been discussed in Ref.\ \cite{Fischer}. Here we will shortly review the essential features.

Let us assume that we have already performed $n-1$ measurements. Our present knowledge about the underlying state of the qubit is then represented by the estimated density operator
\be 
  \rho^{(est)}_{n-1}=\int d\Omega\ w_{n-1}(\theta,\phi)\ \ket{\theta,\phi}\bra{\theta,\phi}
\ee
with a normalized probability density $w_{n-1}$ on the Bloch sphere, i.e., $\int d\Omega\ w_{n-1}(\theta,\phi)=1$. Hence an update of the estimated density operator from one measurement to the next really means an update of its probability density. This can be done in the following way. If we perform a polarization measurement in direction $(\theta_{n},\phi_{n})$ we will find the result $1$ with probability
\be
  P_1(\theta,\phi|\theta_n,\phi_n)=\left|\braket{\theta_n,\phi_n}{\theta,\phi}\right|^2
\ee
and the result $0$ with probability
\be 
  P_0(\theta,\phi|\theta_n,\phi_n)=1-P_1(\theta,\phi|\theta_n,\phi_n)
\ee
conditioned on the general qubit state
\be 
  \ket{\theta,\phi}\equiv\cos\frac{\theta}2\ket1+\sin\frac{\theta}2e^{i\phi}\ket0.
\ee
Therefore, if we measure $i=0$ or $1$ we can update the probability density $w_{n-1}$ according to Bayes rule \cite{Berger} resulting in the probability density 
\be
  w_{n}(\theta,\phi) = {\cal N}P_i(\theta,\phi|\theta_{n},\phi_{n})w_{n-1}(\theta,\phi)
\ee
after $n$ measurements. The normalization constant reads
\be 
  {\cal N}^{-1}=\int d\Omega\ P_i(\theta,\phi|\theta_{n},\phi_{n})w_{n-1}(\theta,\phi).
\ee
From this procedure we clearly see that $w_n$ comprises the results of all $n$ measurements. Since we have no a priori information in the beginning the initial density distribution is given by
\be 
  w_0(\theta,\phi)=\frac1{4\pi}.
\ee

We recall that so far we have still simply postulated a certain measurement direction $(\theta_{n},\phi_{n})$ for the $n$th step. However, the formulation given above allows us to optimize the choice $(\theta_{n},\phi_{n})$ in the sense that we gain as much information as possible out of the $n$th measurement. The expected information gain of the $n$th measurement reads
\be 
  S(\theta_{n},\phi_{n})=-\sum\limits_{i=0}^{1}p^{(est)}_i(\theta_{n},\phi_{n})\ln p^{(est)}_i(\theta_{n},\phi_{n}) 
  \label{entropy}
\ee
using the estimated probabilities
\begin{eqnarray}
  p^{(est)}_1(\theta_{n},\phi_{n})&=&\sandwich{\theta_{n},\phi_{n}}{\rho^{(est)}_{n-1}}{\theta_{n},\phi_{n}} \nonumber\\
  &=&\int d\Omega\ w_{n-1}(\theta,\phi)P_1(\theta,\phi|\theta_{n},\phi_{n})
\end{eqnarray}
and
\be
  p^{(est)}_0(\theta_{n},\phi_{n})=1-p^{(est)}_1(\theta_{n},\phi_{n})
\ee
given by our current knowledge, i.e., given by the density operator $\rho^{(est)}_{n-1}$. The optimized measurement direction then is the one which maximizes $S(\theta_{n},\phi_{n})$, Eq.\ (\ref{entropy}).

Eventually, we will have measured all qubits in this way and hence we arrive at $w_N(\theta,\phi)$. The point $(\theta^{(est)},\phi^{(est)})$ at which $w_N$ takes on its maximum defines the pure state $\ket{\theta^{(est)},\phi^{(est)}}$ which is our final estimate for the original qubit $\ket1_{\vec n}$. The corresponding overlap
\be 
  F=\left|\braket{\theta^{(est)},\phi^{(est)}}{1}_{\vec n}\right|^2
\ee
then measures the fidelity of our estimation.

%--------------------------------------------------------------------------------
\section{Numerical simulations and results}
\label{results}

Using numerical simulations we will now show that the combination of purification and adaptive measurements leads to an improved estimation. Even though we have more and separable qubits at our disposal, if we apply no purification, the case with purification is superior. 

In order to demonstrate this we have calculated the mean fidelity
\be 
  \left<F\right>=\Bigl<\left|\braket{\theta^{(est)},\phi^{(est)}}{1}_{\vec n}\right|^2\Bigr>
\ee
where $\left<\dots\right>$ means choosing $4\times10^4$ initial states $\ket1_{\vec n}$ equally distributed over the Bloch sphere, to ensure that we average over sufficiently many histories of measurement results. In the case of the purification we also have to average over the possible results of the purification process. So for the case of $N$ input qubits all estimations for $M=0, 2, 4, \dots, N$ must be carried out and weighted by the appropriate probabilities $p_M$, Eq.\ (\ref{probm}).

\begin{figure}
\epsfig{file=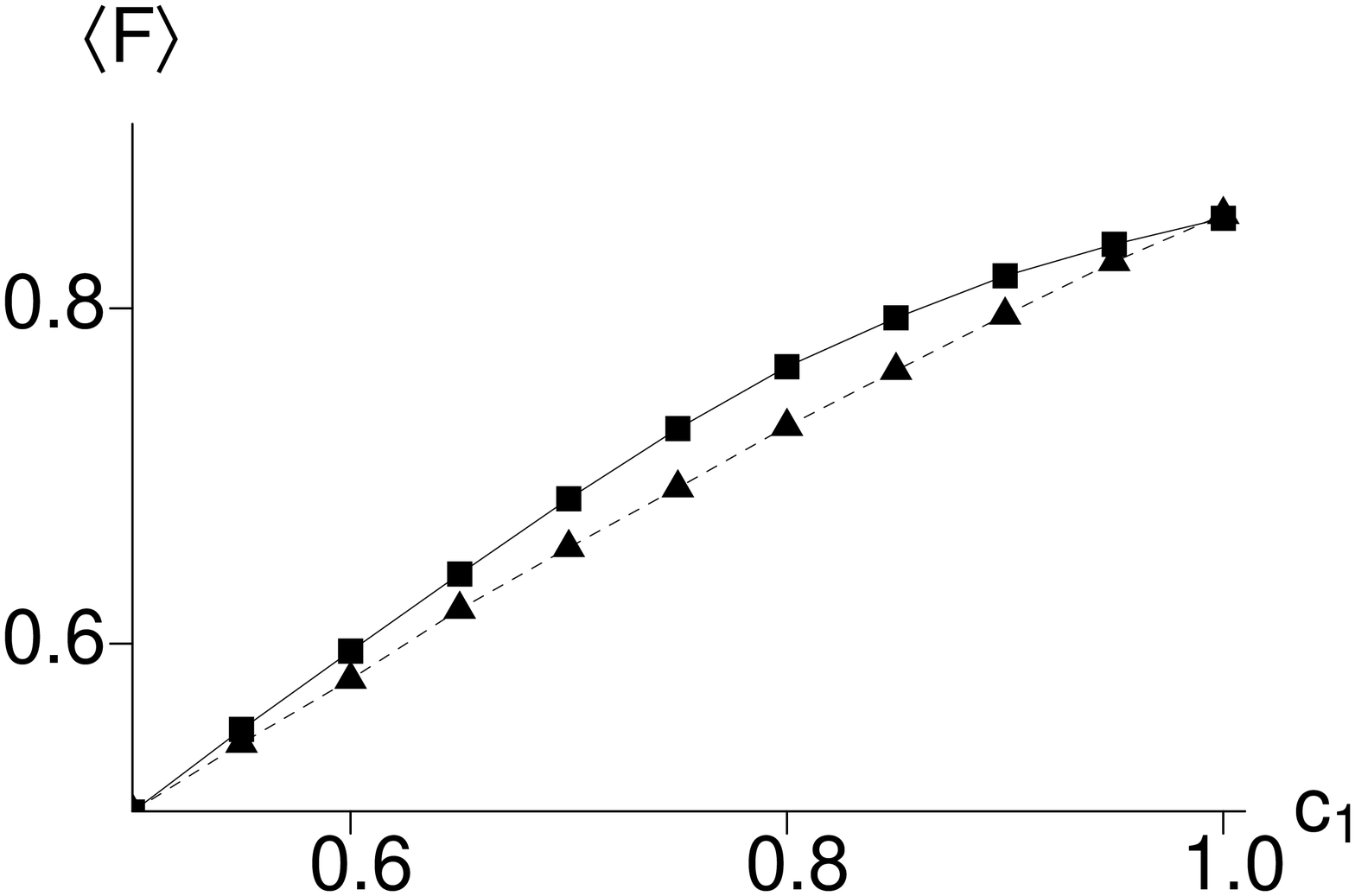,width=80mm}
\caption{Average fidelity $\left<F\right>$ for the estimation of $N=6$ qubits with ({\scriptsize$\blacksquare$}) and without ($\blacktriangle$) purification. Over the whole range from totally mixed states, $c_1=1/2$, to the original pure state, $c_1=1$, we see that purification improves the estimation. Note that for the data points of the unpurified case we always have $6$ qubits at our disposal. The data points involving purification are obtained by estimating $M=0,2,4$ or $6$ purified qubits weighted by the corresponding probability $p_M$, Eq. (\ref{probm}).}
\label{fig1}
\end{figure}

In Fig.\ref{fig1} the average fidelity obtained by the adaptive estimation of $N=6$ purified qubits is compared to the same estimation procedure with unpurified qubits. The results are plotted versus the probability $c_1$ of the initial noisy qubits, Eq.\ (\ref{origrho}). Over the whole range from pure states with $c_1=1$ to totally mixed states with $c_1=1/2$ there is an clear improvement in the estimation results. The relative gain in the fidelity of the measurement  goes up to $5.5\%$ of the unpurified value. Even on the average there is an improvement of about $3.3\%$. So this shows that purification can increase the accessibility of the originally encoded qubit information. Even though the purification protocol leads to entanglement of the purified qubits we find this increase in the average estimation fidelity. That was not clear from the beginning since via this entanglement our single measurements naturally change the quantum state of the qubits not yet measured.

\begin{figure}
\epsfig{file=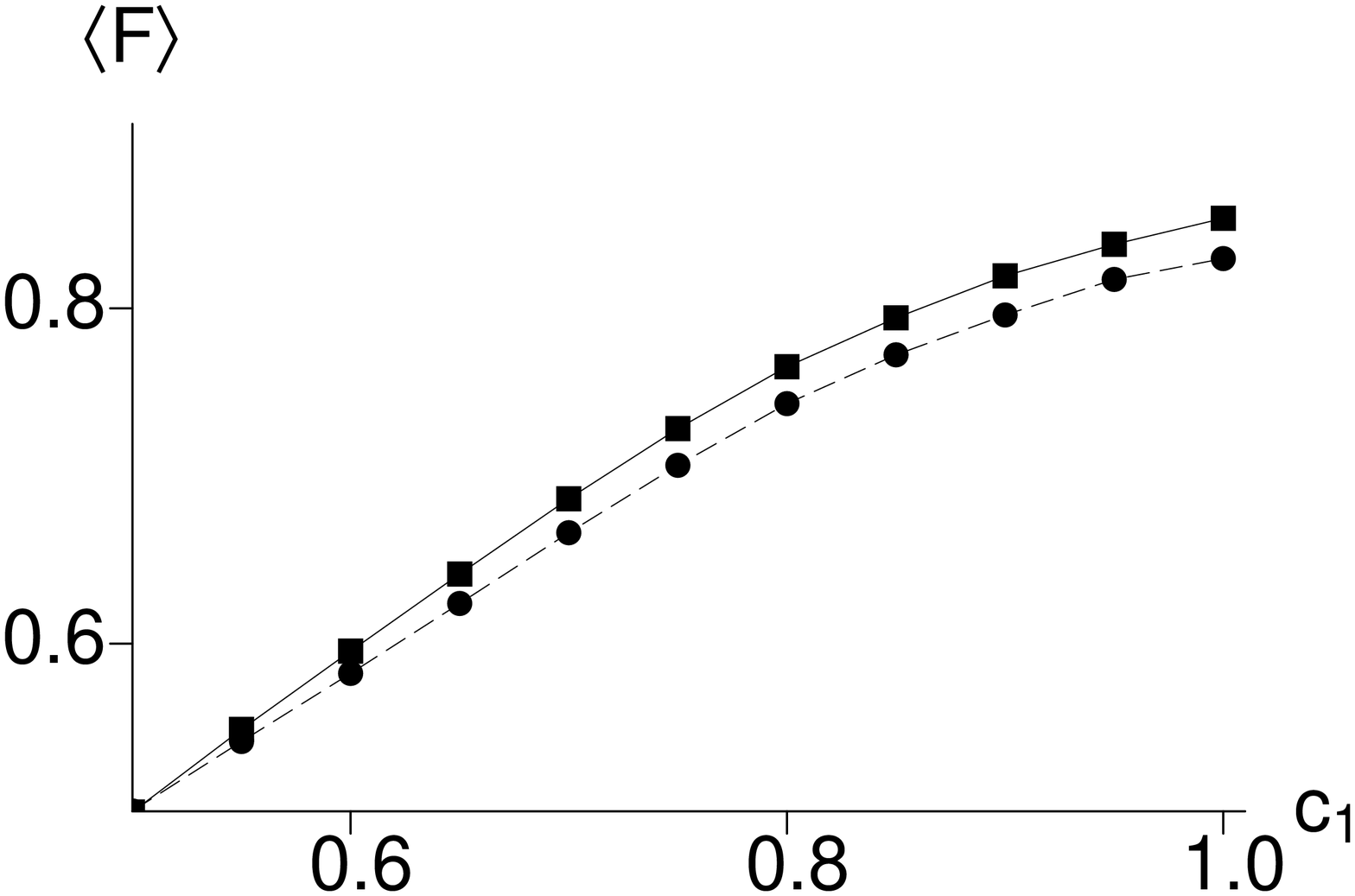,width=80mm}
\caption{Average fidelity $\left<F\right>$ versus the depolarizing parameter $c_1$ for the estimation of $N=6$ purified qubits using a random selection of measurement directions ($\bullet$) and an adaptive strategy ({\scriptsize$\blacksquare$}). If we choose an optimized measurement direction based on all earlier results we clearly get an improved estimation.}
\label{fig2}
\end{figure}

In Fig.\ref{fig2} we compare the adaptive measurement of $N=6$ purified qubits described above to a non-adaptive measurement where the directions $(\theta_{n},\phi_{n})$ of the measurement are choosen randomly on the Bloch sphere. We find that for purified and non-separable qubits one can also improve the estimation quality by using adaptive measurements originally designed for separable states \cite{Fischer}.

\begin{figure}
\epsfig{file=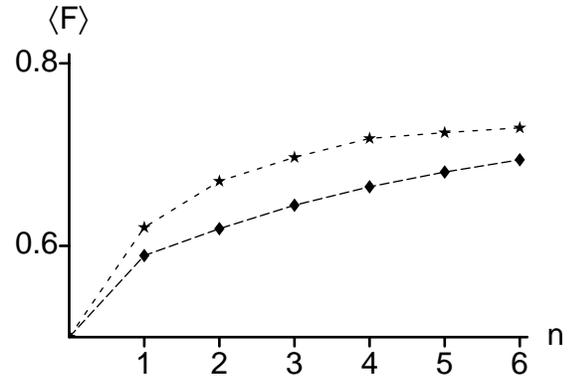,width=80mm}
\caption{Rise of the average fidelity $\left<F\right>$ due to the $n$th measurement for a finite ensemble of $N=6$ qubits depolarized with $c_1=0.75$. If we add a purification step to the estimation procedure ($\bigstar$), we clearly see the improvement compared to the unpurified case ($\blacklozenge$). Note that this improvement stems from the early measurement steps.}
\label{fig3}
\end{figure}

Finally we take a look at the evolution of the estimation fidelity during the estimation process. The plot in Fig.\ref{fig3} shows the fidelity after the $n$-th qubit measurement. Here the initial probability $c_1$ is set to $0.75$. For the unpurified qubits there is a slow increase in this fidelity which keeps up until the last measurement. In contrast to this the fidelity of purified qubits increases only during the first measurements. Then follows a domain of nearly stagnant fidelity. Hence the purification which is basically an increase in the length of the Bloch vector helps the estimation process to rapidly find a good estimation direction. The possible loss of qubits in the purification process does not harm dramatically because of the relatively small gain in the last qubits.

%--------------------------------------------------------------------------------
\section{Conclusions}
\label{conclusions}

Reading of quantum information is a delicate task. The state of a qubit can only be estimated using finite resources. We have shown that adaptive measurements are a useful tool for estimating quantum states. This also holds true for decohered quantum bits which is the more common case in quantum information processing. An advantage of our approach is the reduction to simple spin measurements on single quantum systems which considerably simplifies a possible physical implementation of an estimation procedure.

Qubit purification gives us the possibility to even outrange these results. We have presented results which show that an additional purification step improves the estimation of noisy quantum bits. This means that purification does not only work in the single-qubit case but also increases the accessability of quantum information stored in an ensemble of qubits. This holds true even despite the fact that purification is a probabilistic process and may result in a loss of usable quantum systems.

%--------------------------------------------------------------------------------
\acknowledgements
We would like to thank G. Alber, A. Delgado, and M. Mussinger for helpful discussions. We acknowledge financial support by the DFG Schwerpunktprogramm ``Quanten-Informationsverarbeitung'' and by the European Commission within the IST project QUBITS.

%--------------------------------------------------------------------------------

\newpage

\begin{references}
\bibitem{Footnote1}We still assume that the depolarization is moderate. That is, the original information encoded by Alice using the state $\ket1_{\vec n}$ is not completely destroyed by mixing in the state $\ket0_{\vec n}$. This is why we restrict the depolarizing parameter to $c_1\geq 1/2$.
\bibitem{Footnote2}An odd number of qubits can be treated similarly. To keep the presentation simple we concentrate on the case of an even number $N$.
\end{references}
\end{document}